\documentclass[prl,twocolumn,showpacs,floats]{revtex4}

\usepackage{graphicx}

\begin{document}
\title{Common Fermi Surface Topology and Nodeless Superconducting Gap in K$_{0.68}$Fe$_{1.79}$Se$_2$ and (Tl$_{0.45}$K$_{0.34}$)Fe$_{1.84}$Se$_2$
Superconductors  Revealed from Angle-Resolved Photoemission Spectroscopy}
\author{Lin Zhao$^{1}$, Daixiang Mou$^{1}$,  Shanyu Liu$^{1}$, Xiaowen Jia$^{1}$, Junfeng He$^{1}$, Yingying Peng$^{1}$, Li Yu$^{1}$, Xu Liu$^{1}$, Guodong  Liu$^{1}$, Shaolong He$^{1}$, Xiaoli Dong$^{1}$, Jun Zhang$^{1}$, J. B. He$^{2}$, D. M. Wang$^{2}$, G. F. Chen$^{2}$, J. G. Guo$^{1}$, X. L. Chen$^{1}$,  Xiaoyang Wang$^{3}$, Qinjun Peng$^{3}$, Zhimin Wang$^{3}$, Shenjin Zhang$^{3}$, Feng Yang$^{3}$, Zuyan Xu$^{3}$, Chuangtian Chen$^{3}$  and X. J. Zhou$^{1,*}$}

\affiliation{
\\$^{1}$Beijing National Laboratory for Condensed Matter Physics, Institute of Physics,
Chinese Academy of Sciences, Beijing 100190, China
\\$^{2}$Department of Physics, Renmin University of China, Beijing 100872, China
\\$^{3}$Technical Institute of Physics and Chemistry, Chinese Academy of Sciences, Beijing 100190, China
}
\date{February 5, 2011}
%
%

\begin{abstract}

We carried out high resolution angle-resolved photoemission measurements on the electronic structure and superconducting gap of K$_{0.68}$Fe$_{1.79}$Se$_{2}$ (T$_c$=32 K) and  (Tl$_{0.45}$K$_{0.34}$)Fe$_{1.84}$Se$_2$ (T$_c$=28 K) superconductors. In addition to the electron-like Fermi surface near M($\pi$,$\pi$), two electron-like Fermi pockets are revealed around the zone center $\Gamma$(0,0) in K$_{0.68}$Fe$_{1.79}$Se$_{2}$. This observation makes the Fermi surface topology of K$_{0.68}$Fe$_{1.79}$Se$_{2}$  consistent with
that of  (Tl,Rb)$_x$Fe$_{2-y}$Se$_2$ and (Tl,K)$_x$Fe$_{2-y}$Se$_2$ compounds.  A nearly isotropic superconducting gap ($\Delta$) is observed along the electron-like Fermi pocket near the M point in K$_{0.68}$Fe$_{1.79}$Se$_{2}$ ($\Delta$$\sim$ 9 meV) and (Tl$_{0.45}$K$_{0.34}$)Fe$_{1.84}$Se$_2$ ($\Delta$$\sim$ 8 meV).  The establishment of a universal picture on the Fermi surface topology and superconducting gap in the A$_x$Fe$_{2-y}$Se$_2$ (A=K, Tl, Cs, Rb and etc.) superconductors will provide important information in understanding the superconductivity mechanism of the iron-based superconductors.

\end{abstract}

\pacs{74.70.-b, 74.25.Jb, 79.60.-i, 71.20.-b}

\maketitle

\newpage

The latest discovery of superconductivity with a T$_c$ above 30 K in a new A$_x$Fe$_{2-y}$Se$_2$ (A=K, Tl, Cs, Rb and etc.) system\cite{JGGuo,Switzerland,Mizuguchi,MHFang,GFChen} has triggered a new wave of broad interest in the iron-based high temperature superconductors\cite{Kamihara,ZARenSm,RotterSC,MKWu11,CQJin111}.  A couple of unique characteristics of the A$_x$Fe$_{2-y}$Se$_2$ system provide new perspectives that ask for rethinking and re-examination of ideas which have been proposed for other iron-based superconductors, such as the effect of Fe vacancy and structural modulation on superconductivity\cite{MHFang,GFChen,ZWang,PZavalij}, the nature of the underlying parent compound\cite{MHFang,GFChen,QMSi,YiZhou}, the role of electron scattering across the bands between the zone center $\Gamma$(0,0) and zone corner M($\pi$,$\pi$) on superconductivity, and the pairing symmetry of this new system with a distinct Fermi surface topology\cite{YiZhou,FWang}.  Band structure calculations of A$_x$Fe$_{2-y}$Se$_2$\cite{LJZhang,IRShein,XWYan} suggest that the large electron doping in this system leads to the disappearance of the hole-like Fermi surface pockets around the $\Gamma$ point that are commonly present in other Fe-based compounds.  In this case, the peculiar Fermi surface topology near $\Gamma$ in the A$_x$Fe$_{2-y}$Se$_2$ superconductors would make it unlikely to have electron scatterings from the hole-like bands near $\Gamma$ to the electron-like bands near M that are considered to play an important role in the electron pairing and superconductivity in the Fe-based superconductors by some theories\cite{Kuroki,FeSCMagnetic}.  Experimental investigations on the electronic structure and the superconducting gap of A$_x$Fe$_{2-y}$Se$_2$ superconductors are thus crucial for understanding the physical properties and  the pairing mechanism in the iron-based superconductors.

\begin{figure}[tbp]
\begin{center}
\includegraphics[width=1.00\columnwidth,angle=0]{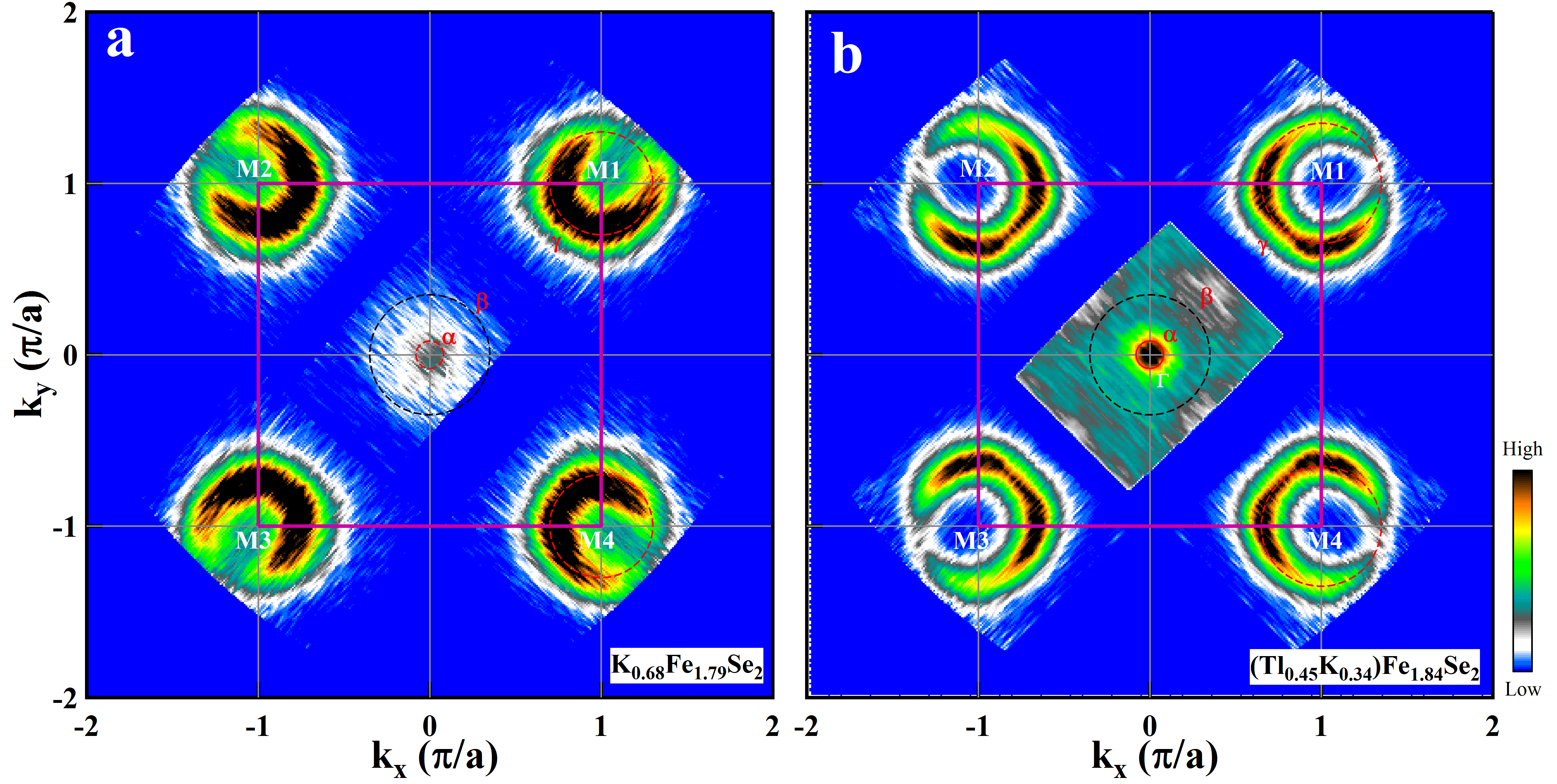}
\end{center}
\caption{Fermi surface mapping of K$_{0.68}$Fe$_{1.79}$Se$_{2}$ superconductor (T$_c$=32 K)(a)  and (Tl$_{0.45}$K$_{0.34}$)Fe$_{1.84}$Se$_2$ superconductor (T$_c$=28 K) (b) measured by using h$\nu$=21.2 eV light source.  Near the M($\pi$,$\pi$) point, one Fermi surface sheet is clearly observed which is marked as $\gamma$ (for the sake of clarity, we refer the four equivalent M points in the first BZ as M1, M2, M3 and M4). Near the $\Gamma$(0,0) point, in addition to a tiny Fermi pocket observed which is marked as $\alpha$, a weak large Fermi surface sheet (marked as $\beta$) is also discernable.
}
\end{figure}

\begin{figure*}[floatfix]
\begin{center}
\includegraphics[width=1.80\columnwidth,angle=0]{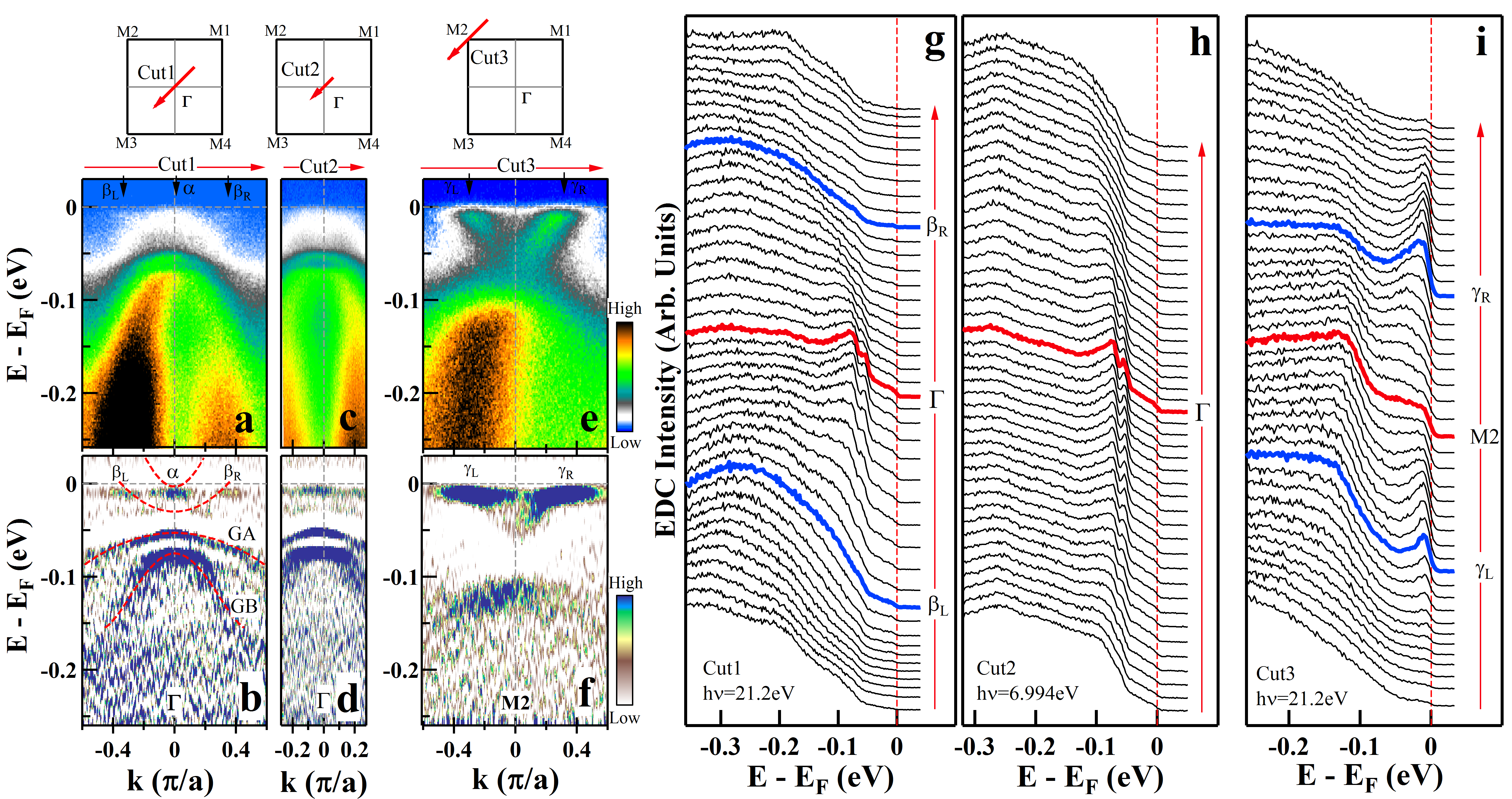}
\end{center}
\caption{Band structure and photoemission spectra of K$_{0.68}$Fe$_{1.79}$Se$_{2}$ measured along typical high symmetry cuts.
(a). Band structure along the Cut 1 crossing the $\Gamma$ point measured by using h$\nu$=21.2 eV light source; the location of the cut is shown on the top of Fig. 2a.  (b). Corresponding EDC second derivative image of Fig. 2a. The $\alpha$ band and two Fermi crossings of the $\beta$ band ($\beta_L$ and $\beta_R$) are marked. Two inverse-parabolic GA and GB bands are also marked.  (c). Band structure along the Cut 2 crossing the $\Gamma$ point measured by using h$\nu$=6.994 eV VUV laser.  (d). Corresponding EDC second derivative image of Fig. 2c. (e). Band structure along the Cut 3 crossing the M2 point measured by using h$\nu$=21.2 eV.  (f). Corresponding EDC second derivative image of Fig. 2e. Two Fermi crossings of the $\gamma$ band ($\gamma_L$ and $\gamma_R$) are marked. The photoemission spectra (EDCs) corresponding to the Cut1, Cut2 and Cut3 are shown in (g), (h) and (i), respectively.
}
\end{figure*}

Angle-resolved photoemission spectroscopy (ARPES) is a powerful tool to directly measure the electronic structure and superconducting gap of superconductors\cite{Damascelli}.  Some initial ARPES measurements on K$_x$Fe$_{2-y}$Se$_2$ did not observe Fermi surface near $\Gamma$\cite{TQian} or observed only a trace of a tiny electron-like pocket near $\Gamma$\cite{YZhang}. These seem to be in agreement with the band structure calculations\cite{LJZhang,IRShein,XWYan}. However, in the ARPES measurement on (Tl,Rb)$_x$Fe$_{2-y}$Se$_2$\cite{DXMou}, two electron-like Fermi surface sheets are observed near $\Gamma$, with the large one  having a similar size as the one near the electron-like pocket around M. The existence of two electron-like pockets near $\Gamma$ is also reported in (Tl,K)$_x$Fe$_{2-y}$Se$_2$\cite{XPWang}. These results raise an obvious issue on whether the Fermi surface topology of K$_x$Fe$_{2-y}$Se$_2$ is different from (Tl,Rb,K)$_x$Fe$_{2-y}$Se$_2$; the resolving of this issue is important for sorting out general electronic structure features in understanding the Fe-based superconductors.

In this paper, we report the observation of two electron-like Fermi surface sheets  around the zone center $\Gamma$(0,0) in K$_{0.68}$Fe$_{1.79}$Se$_{2}$ superconductor (T$_c$=32 K) revealed from our high resolution ARPES measurements.  This is different from the previous ARPES reports that no Fermi pocket or only one tiny Fermi pocket is present near $\Gamma$ in K$_{x}$Fe$_{2-y}$Se$_2$\cite{TQian,YZhang}. The observation of two electron-like Fermi pockets near $\Gamma$  makes the Fermi surface topology of K$_{x}$Fe$_{2-y}$Se$_{2}$  consistent with that in (Tl,Rb)$_x$Fe$_{2-y}$Se$_2$\cite{DXMou} and (Tl,K)$_x$Fe$_{2-y}$Se$_2$\cite{XPWang}, thus establishing a coherent picture of Fermi surface topology in the A$_x$Fe$_{2-y}$Se$_2$ (A=K, Tl, Cs, Rb and etc.) system.   We observe nearly isotropic superconducting gap ($\Delta$) around the Fermi pocket near M in K$_{0.68}$Fe$_{1.79}$Se$_{2}$ ($\Delta$$\sim$ 9 meV) and (Tl$_{0.45}$K$_{0.34}$)Fe$_{1.84}$Se$_2$ ($\Delta$$\sim$ 8 meV). The general picture on the Fermi surface topology and its associated superconducting gap in the A$_x$Fe$_{2-y}$Se$_2$ (A=K, Tl, Cs, Rb and etc.) superconductors will provide key insights in understanding the iron-based superconductors.

High resolution angle-resolved photoemission (ARPES) measurements were carried out by using our lab system equipped with a Scienta R4000 electron energy analyzer\cite{GDLiu}. We used Helium discharge lamp as the light source which provides photons with an energy of h$\upsilon$= 21.218 eV (Helium I), as well as vacuum ultraviolet (VUV) laser which provides h$\upsilon$= 6.994 eV photons.  The energy resolution was set at 10 meV for the Fermi surface mapping (Fig. 1) and band structure measurements (Fig. 2) and at 4 meV for the superconducting gap measurements (Figs. 3 and 4). The angular resolution is $\sim$0.3 degree. The Fermi level is referenced by measuring on a clean polycrystalline gold that is electrically connected to the sample.  The K$_{0.68}$Fe$_{1.79}$Se$_{2}$ and (Tl$_{0.45}$K$_{0.34}$)Fe$_{1.84}$Se$_2$ single crystals were grown by the Bridgeman method\cite{GFChen}.  The composition of the crystals were  analyzed by the energy dispersive X-ray (EDX) spectroscopy.  Electrical resistivity and  DC magnetic susceptibility measurements show that the crystals exhibit a  sharp superconducting transition at T$_c$$\sim$32 K (transition width of $\sim$1 K) for K$_{0.68}$Fe$_{1.79}$Se$_{2}$ and T$_c$$\sim$28 K (transition width of $\sim$1 K) for (Tl$_{0.45}$K$_{0.34}$)Fe$_{1.84}$Se$_2$. The crystal was cleaved {\it in situ} and measured in vacuum with a base pressure better than 5$\times$10$^{-11}$ Torr.

Fig. 1 shows Fermi surface mapping of K$_{0.68}$Fe$_{1.79}$Se$_{2}$ (Fig. 1a) and (Tl$_{0.45}$K$_{0.34}$)Fe$_{1.84}$Se$_2$ (Fig. 1b) superconductors.  The band structure of K$_{0.68}$Fe$_{1.79}$Se$_{2}$ along two typical high symmetry cuts are shown in Fig. 2.   An electron-like Fermi surface is clearly observed around M($\pi$,$\pi$), similar to previous ARPES results on K$_{x}$Fe$_{2-y}$Se$_{2}$\cite{TQian,YZhang}, (Tl,Rb)$_x$Fe$_{2-y}$Se$_2$\cite{DXMou} and (Tl,K)$_x$Fe$_{2-y}$Se$_2$\cite{XPWang}.  Near the $\Gamma$ point, a tiny Fermi pocket (denoted as $\alpha$) is obvious which is possibly formed by an electron-like band with its bottom nearly touching the Fermi level.  In addition, one can observe a rather weak but discernable electron-like Fermi surface sheet (denoted as $\beta$) near $\Gamma$ in both K$_{0.68}$Fe$_{1.79}$Se$_{2}$ (Fig. 1a) and (Tl$_{0.45}$K$_{0.34}$)Fe$_{1.84}$Se$_2$ (Fig. 1b), with its size being similar to that of the electron-like pocket near M.

\begin{figure}[tbp]
\begin{center}
\includegraphics[width=0.96\columnwidth,angle=0]{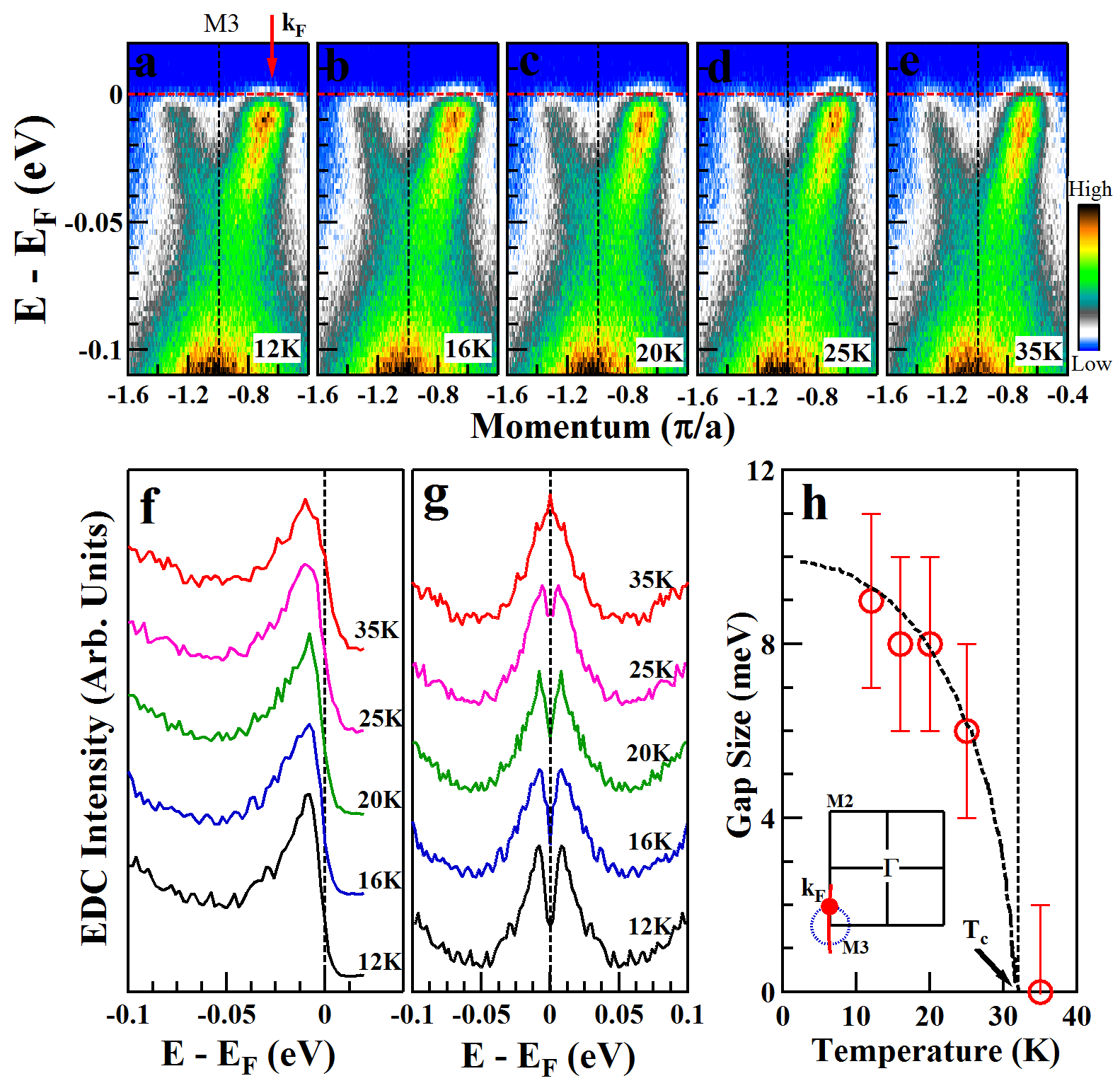}
\end{center}
\caption{Temperature dependence of the superconducting gap of K$_{0.68}$Fe$_{1.79}$Se$_{2}$ (T$_c$$\sim$32 K) along the $\gamma$ Fermi pocket near M.  (a-e) show photoemission images taken at different temperatures along a cut near M3; the location of the cut is marked in the bottom-left inset of (h).  (f). Photoemission spectra measured at different temperatures at the Fermi crossing k$_F$ of the $\gamma$ band, as marked in (a).  (g). The corresponding symmetrized EDCs of (f).  (h). Temperature dependence of the measured superconducting gap (empty red circles). The black dashed line is a curve following the BCS form.
}
\end{figure}

The existence of the $\beta$ Fermi pocket near the $\Gamma$ point in K$_{0.68}$Fe$_{1.79}$Se$_{2}$ can also be identified from the measured band structure (Fig. 2a and Fig. 2b).  We note that the feature of the $\beta$ band (Fig. 2a) and its associated Fermi surface (Fig. 1a)  near $\Gamma$ are rather weak in K$_{0.68}$Fe$_{1.79}$Se$_{2}$, much weaker than in (Tl,Rb)$_x$Fe$_{2-y}$Se$_2$\cite{DXMou}; this is probably why it was not revealed before\cite{TQian,YZhang}.  We also notice that the band structure of K$_{0.68}$Fe$_{1.79}$Se$_{2}$ near the $\Gamma$ point (Figs. 2a, 2b, 2c and 2d) presents some new features that were not observed before.  As shown in Fig. 2b, in addition to the electron-like $\alpha$ band and the  electron-like $\beta$ band, at least two more bands are clearly present within the measured energy window. The observation of the hole-like GB band is consistent with other measurements on K$_{x}$Fe$_{2-y}$Se$_{2}$\cite{TQian,YZhang} that is also commonly observed in (Tl,Rb)$_x$Fe$_{2-y}$Se$_2$\cite{DXMou} and (Tl,K)$_x$Fe$_{2-y}$Se$_2$\cite{XPWang}.  However, the presence of a new GA band is very clear in our measurement (Figs. 2b and 2d) which was not observed in the previous measurements\cite{TQian,YZhang}. The revelation of this GA band is important when comparing the experimental results with the band structure calculations and considering electron scatterings between various bands.

The observation of two electron-like Fermi pockets, $\alpha$ and $\beta$, around $\Gamma$ in K$_{0.68}$Fe$_{1.79}$Se$_{2}$ is interesting.  It is distinct from other Fe-based compounds where hole-like Fermi surface sheets are expected around the $\Gamma$ point\cite{DJSingh1111,Kuroki}. It is also different from band structure calculations\cite{LJZhang,IRShein,XWYan,YZhang,TQian} and previous ARPES measurements\cite{YZhang,TQian} on K$_x$Fe$_{2-y}$Se$_2$ that only suggest disappearance of hole-like Fermi surface sheets near the $\Gamma$ point. It becomes now consistent with the ARPES measurements on (Tl,Rb)$_x$Fe$_{2-y}$Se$_2$\cite{DXMou} and (Tl,K)$_x$Fe$_{2-y}$Se$_2$\cite{XPWang} to provide a general picture on the Fermi surface topology in the A$_x$Fe$_{2-y}$Se$_2$ (A=K, Tl, Cs, Rb and etc.) superconductors.

Now we turn to investigate the superconducting gap in the K$_{0.68}$Fe$_{1.79}$Se$_{2}$ and (Tl$_{0.45}$K$_{0.34}$)Fe$_{1.84}$Se$_2$ superconductors. Since the $\beta$ feature near $\Gamma$ is too weak to give reasonable information on the superconducting gap, we will focus in this paper on the superconducting gap along the $\gamma$ Fermi surface near M.  Figs. 3(a-e) show the photoemission images measured on K$_{0.68}$Fe$_{1.79}$Se$_{2}$
along a cut near M (its location shown in the bottom-left inset of Fig. 3h) at different temperatures. The photoemission spectra (energy distribution curves, EDCs) on the Fermi momentum at different temperatures are shown in Fig. 3f.  To visually inspect possible gap opening and remove the effect of Fermi distribution function near the Fermi level, these original EDCs are symmetrized to get spectra in Fig. 3g, following the procedure that is commonly used in high temperature cuprate superconductors\cite{MNorman}.  As seen from Fig. 3g, there is a clear superconducting gap opening below T$_c$$\sim$ 32 K which is closed above T$_c$.  The superconducting gap size is extracted from the peak position of the symmetrized EDCs in this paper\cite{MNorman} (Fig. 3g); it is $\sim$9 meV at 12 K and its temperature dependence roughly follows the BCS-type form (Fig. 3h).

In order to measure the momentum-dependence of the superconducting gap,  we took high resolution Fermi surface mapping of  the $\gamma$ Fermi pocket at M  for K$_{0.68}$Fe$_{1.79}$Se$_{2}$ (Fig. 4a)  and (Tl$_{0.45}$K$_{0.34}$)Fe$_{1.84}$Se$_2$ (Fig. 4e) superconductors.  Fig. 4b shows photoemission spectra around the $\gamma$ Fermi pocket (Fig. 4a) measured in the superconducting state (T= 15 K); the corresponding symmetrized photoemission spectra are shown in Fig. 4c.  The  superconducting gap (Fig. 4d), extracted by picking up the peak position of the symmetrized EDCs (Fig. 4c),  is nearly isotropic with a size of (9$\pm$2) meV.    By the same procedure, the  superconducting gap around the $\gamma$ Fermi pocket near M for the (Tl$_{0.45}$K$_{0.34}$)Fe$_{1.84}$Se$_2$ superconductor (Fig. 4h) is also nearly isotropic with a size of (8$\pm$2) meV.

\begin{figure}[t]
\begin{center}
\includegraphics[width=1.00\columnwidth,angle=0]{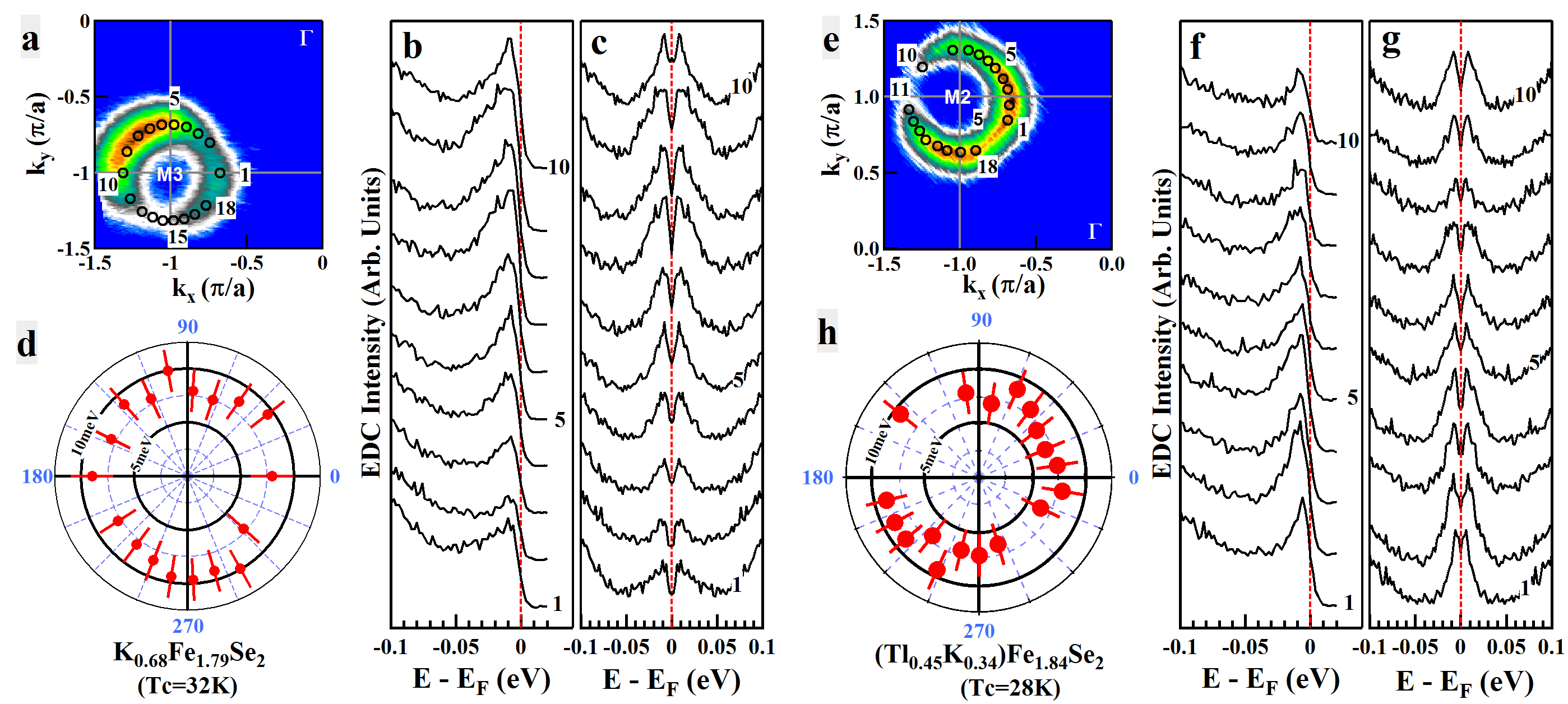}
\end{center}
\caption{Momentum dependent superconducting gap of K$_{0.68}$Fe$_{1.79}$Se$_{2}$ superconductor (T$_c$=32 K)  and (Tl$_{0.45}$K$_{0.34}$)Fe$_{1.84}$Se$_2$ superconductor (T$_c$=28 K) measured along the $\gamma$ Fermi surface sheet near M at a temperature of 15 K. (a). High resolution Fermi surface mapping of K$_{0.68}$Fe$_{1.79}$Se$_{2}$ near M3;  the corresponding Fermi crossings are marked by empty black circles. (b) and (c) show several typical EDCs along the $\gamma$ Fermi surface and their corresponding symmetrized EDCs, respectively.  (d). Momentum dependence of the superconducting gap along the $\gamma$ Fermi surface sheet (solid red circles).  (e), (f), (g) and (h) show, respectively,  the high resolution Fermi surface mapping near M2, EDCs along the Fermi surface, their corresponding symmetrized EDCs, and the obtained momentum-dependent superconudtcing gap for the (Tl$_{0.45}$K$_{0.34}$)Fe$_{1.84}$Se$_2$ superconductor.
}
\end{figure}

In summary, we have identified two electron-like Fermi pockets near the $\Gamma$ point in K$_{0.68}$Fe$_{1.79}$Se$_{2}$ and (Tl$_{0.45}$K$_{0.34}$)Fe$_{1.84}$Se$_2$ superconductors. This has established a consistent picture on the Fermi surface topology in the A$_x$Fe$_{2-y}$Se$_2$ (A=K, Tl, Cs, Rb and etc.) superconductors.  The distinct Fermi surface topology in the A$_x$Fe$_{2-y}$Se$_2$  superconductors definitely asks for re-evaluation of the pairing mechanisms, based on electron scatterings between the bands near $\Gamma$ and the bands near M,  proposed before for other Fe-based superconductors \cite{Kuroki,FeSCMagnetic}.   We have observed nearly isotropic superconducting gap around the $\gamma$ Fermi pocket near the M point in K$_{0.68}$Fe$_{1.79}$Se$_{2}$ ($\Delta$$\sim$ 9 meV) and (Tl$_{0.45}$K$_{0.34}$)Fe$_{1.84}$Se$_2$ ($\Delta$$\sim$ 8 meV). These are consistent with other ARPES measurements\cite{YZhang,DXMou,XPWang} to build a general picture on an isotropic superconducting gap along the $\gamma$ Fermi surface near M.   These results, together with the observation of nearly isotropic superconducting gap along the $\beta$ pocket near $\Gamma$\cite{DXMou,XPWang}, indicate that the A$_x$Fe$_{2-y}$Se$_2$ superconductors are nodeless in its gap structure, a fact that appears to favor an s-wave symmetry or a nodeless d-wave symmetry\cite{YiZhou,FWang}.   These rich information on the Fermi surface topology and the associated superconducting gap will provide crucial information and constraints on understanding the superconductivity mechanism in  the  Fe-based superconductors.

XJZ thanks the funding support from NSFC (Grant No. 10734120) and the MOST of China (973 program No: 2011CB921703).

$^{*}$Corresponding author: XJZhou@aphy.iphy.ac.cn


\end{document}